\documentclass[12pt,aps,bm,epsfig,amssymb]{article}
\usepackage{epsfig,amsmath,amsfonts,float}
\newcommand{\R}[0]{{\mathbb{R}}}
\newcommand{\Z}[0]{{\mathbb{Z}}}
\newcommand{\cH}{{\cal H}}

\newcommand{\cX}{{\cal X}}
\newcommand{\cI}{{\cal I}}
\def\C{{\mathbb{C}}}
\def\Z{{\mathbb{Z}}}
\def\R{{\mathbb{R}}}

\newtheorem{Lemma}{Lemma}
\newtheorem{Definition}{Definition}
\newtheorem{Theorem}{Theorem}

\newcommand{\bef}{\begin{figure}}
\newcommand{\eef}{\end{figure}}

\newcommand{\bei}{\begin{itemize}}
\newcommand{\eei}{\end{itemize}}
\newcommand{\bea}{\begin{eqnarray}}

\newcommand{\eea}{\end{eqnarray}}
\newcommand{\bequ}{\begin{equation}}
\newcommand{\eequ}{\end{equation}}

\begin{document}


\title
{A Quantum Broadcasting Problem in Classical Low Power
Signal  Processing}

\author{Dominik Janzing and Bastian Steudel\\
[1ex]
{\small Institut f\"{u}r Algorithmen und Koginitive Systeme,} 
\\ 
{\small Universit\"{a}t Karlsruhe,} \\
{\small Am Fasanengarten 5,}\\ {\small 76 131 Karlsruhe, Germany}
}

\date{September 21, 2006}

\maketitle 

\begin{abstract}
We pose a problem called ``broadcasting Holevo-information'':\\
given an unknown state taken from an ensemble, the task is
to generate a bipartite state transfering as much Holevo-information
to each copy as possible.  

We argue that upper bounds 
on the average information over both copies imply
lower bounds on the quantum capacity required to send
the ensemble without information loss. This is because a 
channel with zero quantum capacity has a unitary extension 
transfering at least as much information to its environment as it transfers 
to the output.

For an ensemble being the time orbit of a pure state 
under a Hamiltonian evolution, we
derive 
such a bound on 
 the required quantum capacity
in terms of 
properties of the input and  
output energy distribution. 
Moreover, we discuss relations between the
broadcasting problem and entropy power inequalities. 

The broadcasting problem arises when a signal should be 
transmitted by a time-invariant device such that
the outgoing signal has the same timing information as the 
incoming signal had. Based on previous results we argue that this 
establishes a link between
quantum information
theory and the theory of low power computing because
the loss  of timing information
implies loss of free energy.




\end{abstract}

\section{Introduction}

Quantum information theory and the theory of low-power processing
are currently quite different scientific disciplines.
Even though future low power computers will operate 
more and more on the nanoscale and therefore in the quantum regime
(e.g. single electron transistors, spintronic networks \cite{Awschalom}),
superpositions of logically different states being 
crucial for quantum computing \cite{NC}, are not intended to 
occur in low-power computing devices. 

On the other hand, quantum computing research is little interested
in issues of low power processing. The control of quantum systems
involves large laboratory equipment and even 
power consumption
rates for logical operations that are in the magnitude of
usual classical chips seem currently to be out of reach. 

To understand limitations of low-power information processing 
it is useful 
to construct theoretical models of computers 
which process information without
consuming energy, i.e., the process is implemented in an energetically 
closed physical system.
In our opinion, 
discussions on
fundamental issues like 
bounds on power consumption require   a {\it quantum} theoretical
description. 
Interesting quantum models of computers 
 being closed physical
systems can be found in Refs.~\cite{Benioff,Feynman:85,Margolus:90,Ergodic,ErgodicQutrits}. 
Remarkably, it is common to all these models that the synchronization
is based upon 
some propagating wave or particle and that the quantum uncertainty of its 
position leads to an ill-defined logical state of the computer.
In other words, the clock is entangled with the data register.
It seems as if the clocking issue brings some aspects of 
{\it quantum} information theory into the field of low power computing.
This is not surprising for the following reason:
the states of a 
quantum mechanical system have a consistent {\it classical} description
only if the attention is restricted to a set of mutually commuting
density matrices. But the Hamiltonian dynamics  
automatically generated non-commuting density matrices from a given one. 
Hence the dynamical aspect makes it necessary to include
quantum superpositions into the description. 
Note that this is also
in the spirit of Hardy's paper 
``Quantum  theory from five reasonable axioms'' 
\cite{Hardy}, saying that every statistical theory that satisfies some
very natural axioms is quantum, as soon as it makes {\it 
continuous reversible} dynamical evolution possible. 

If signal propagation in future low-power devices takes place
in a system being (approximately) thermodynamically  closed
it must be described by a quantum Hamiltonian evolution. The idea
of this article is that processing such signals
leads to quantum broadcasting problems for two reasons:

First, it is a natural problem to {\it distribute} signals
(like clock signals)
 to several 
devices. The timing information carried by a signal whose quantum state
is a density operator within  a family of non-commuting
states cannot be considered as classical information, its
distribution is therefore some kind of broadcasting problem.
The results in \cite{clock,viva2002} indicate that
no-broadcasting theorems are expected to get
relevant for the distribution 
when the signal energy is reduced to a scale where
quantum energy-time uncertainty becomes the limiting factor for
the accuracy of clocking. 

The second reason why broadcasting problems come into play 
is more subtle. If such a clocking signal enters a device and
triggers the transmission of an output signal we may
desire  that the output should have as much
timing information as the input (in a sense that will be
further specified later). Whether channels with zero quantum capacity
are able to satisfy this requirement 
is a question that is linked to
quantum broadcasting. 

The intention of this article is to describe a special kind of
broadcasting problem. 
In contrast to the usual setting \cite{NoBroadcast}, the task is not
to obtain output states that are close to the inputs.
The problem is to broadcast the
{\it Holevo-information} of an ensemble of non-commuting quantum density 
matrices such that each party gets almost the same amount 
of Holevo-information as
the original ensemble possessed. The use of entropy-like 
information measures makes  it possible  to draw
connections to thermodynamics.

In this paper, the ensemble of
non-commuting states will always be given by the Hamiltonian 
time evolution of a given state. Even though the problem 
of broadcasting Holevo-information makes also sense for general ensembles,
time evolution is the most obvious way how 
non-commuting
ensembles occur in devices that are not designed to do 
quantum information processing. 

It seems to be hard to derive general bounds on the information loss
of each copy when the Holevo-information 
is broadcast. Thus, we will  only conjecture that it is not possible 
for non-commuting ensembles to get full Holevo-information for both
copies.
The intention of this article is therefore 
rather to pose  the broadcasting problem  
and show its relevance than to solve it.
However, for pure input states 
we will give one lower bound on the loss
that depends on the energy distribution of input and output signals.


The paper is organized as follows. In Section~\ref{TimeIn} we introduce
time-invariant signal processing devices
and argue that in this setting timing information is  a resource that
can never be increased. In Section~\ref{Broad} 
we formally state the problem of broadcasting Holevo-information 
in the general case and in the case of timing information.
In Section~\ref{PureProd} we argue that the broadcasting problem
leads to the question how the 
Holevo-information of an ensemble of bipartite states 
is related  to the information of the ensembles of the corresponding 
reduced states.
We discuss this information deficit for
the special case of pure product states
where the problem is 
related to the entropy power inequalities of classical information theory. 
In Section~\ref{PureEnt} 
we
derive a 
bound on the information deficit
 that depends
on the energy spectral measure of input and output signal. 
In Section~\ref{Class} we show that the results
imply lower bounds on the quantum  capacity required for 
 lossless
transmission of signals having small energy uncertainty 
in a time-covariant way. Section~\ref{Sec:Thermo}
derives lower bounds on the loss of free energy implied by the loss
of timing information caused by a channel with too little quantum capacity. 
This describes an even tighter link between quantum information theory 
and the theory of low-power signal processing.

\section{Quantum model of time-invariant signal processing devices}

\label{TimeIn}

As already stated,
the problem of transmitting non-commuting ensembles of quantum states
arises most naturally for ensembles that are given by the 
Hamiltonian time evolution
of a given state. 
Such an ensemble may, for instance, describe 
the density matrix of a propagating signal before or after it is processed
by the device.
If all clocking signals that enter a given device are included 
into the formal description, the quantum operation mapping the input
onto the output 
is {\it time-invariant}.
As we will describe below, such a device
cannot increase the timing information. The latter 
is therefore considered as a resource. The idea 
that devices with non-zero 
quantum capacity  seem to deal with this resource more carefully 
than classical channels is essential for this article. 

Now we introduce the abstract description of time-invariant devices.
Here a device may be a transistor, an optical element or some other
system with input and output signals. 
The signal may, 
for instance, be an electric pulse, a light pulse, or an 
acoustical signal.
We consider it as a physical system with some Hilbert space $\cH$
and the state is a density operator $\rho$ acting on $\cH$.
For the examples mentioned above,
the space $\cH$ will typically be infinite dimensional
since one may e.g. think of position degrees of freedom that are encoded
into $\rho$.
Before and after the signal is processed in the device
its free time evolution 
is generated by a Hamiltonian
$H$ (i.e. a densely defined self-adjoint operator on $\cH$) 
and reads
\begin{equation}\label{Motion}
\alpha_t(\rho):=e^{-iHt} \rho\, e^{iHt}\,.
\end{equation}
We assume that
input state $\rho$ and its output $G(\rho)$ are related by 
some completely positive trace-preserving map $G$ satisfying
the covariance condition
\begin{equation}\label{TCov}
G(\alpha_t(\rho))=\alpha_t(G(\rho)) \hspace{2cm} \forall \rho\,.
\end{equation}
There are rather different situations where the covariance
condition is satisfied. One example would be if the 
interactions between signal and  device are weak. 
A more interesting justification is the following.
Consider a signal propagating towards the device by its own
autonomous Hamiltonian time evolution until it begins to interact with
the latter.
Then it leaves the device (as a possibly modified signal) 
and as soon as the interaction with the device is negligible
it is again subjected to its Hamiltonian only. 
Such a process
should be considered as a quantum stochastic
analogue of a scattering process (see \cite{TimeCovariant} for details) and 
the time covariance condition (\ref{TCov}) 
is then a generalization of
the statement that
the $S$-``matrix'' of a scattering process commutes with the free Hamiltonian
evolution of the incoming and outgoing particle \cite{LP67}.
Note that the existence of an unitary $S$ operator would require 
devices which  preserve the purity of the input. 

In \cite{TimeCovariant} we have given a quite explicit description
of the set of CP maps satisfying this covariance condition.
Here it is more interesting to discuss the implications
of  covariance. We first rephrase the definition of timing information
used in \cite{Referenz} (see also \cite{GroupCovariantThermo} for
a more general context).

Recall that the Holevo-information of an ensemble of quantum states
$\rho_x$ with probability measure $p$ 
(denoted by $\{p(x),\rho_x\}_x$) 
is defined by \cite{NC}
\[
\cI:=S\Big(\int  \rho_x  \,d p(x) \Big)-\int S(\rho_x) \,dp(x)\,,
\] 
where the measure-theoretic integral reduces to sums when $p$ is
supported by a countable set of points.  
Here 
\begin{equation}
S(\rho)=-tr(\rho \log \rho)
\end{equation}
is the von-Neumann entropy and
the base of the logarithm remains unspecified.
In the sequel we will measure entropy 
in bits or nats since sometimes one unit is more natural
and sometimes the other. 

Timing information refers to a  specific  ensemble, namely the
orbit with respect to a unitary one-parameter group:

\begin{Definition}[Timing Information]${}$\\ \label{Tim}
Let $\rho$ be the state of a quantum system whose Hamiltonian $H$
has discrete spectrum. Then its timing information 
is defined as 
\begin{equation}\label{EqTim}
\cI:=S(\overline{\rho})-S(\rho)\,,
\end{equation}
where $\overline{\rho}$ denotes the time average
\[
\overline{\rho}:=\lim_{T\to\infty}
\frac{1}{T}\int_0^T \alpha_t(\rho) dt =\sum_x R_x \rho R_x \,,
\]
 $\alpha_t$ is defined as in eq.~(\ref{Motion}) and $(R_x)$ 
is the family of spectral projections (with eigenvalues $x$ 
corresponding to the system Hamiltonian. For pure states 
$\rho=|\psi\rangle \langle \psi|$ we have $S(\rho)=0$.
Thus, $\cI$ is the entropy of $\overline{\rho}$ which is 
then exactly the entropy of a classical random variable $X$ describing
the distribution of energy values with
$P(X=x):=\langle \psi |R_x|\psi\rangle$.    
\end{Definition}

\noindent
Note that it is a well-known question
to what extent 
information on  reference frames in time or space requires
quantum communication or profits from it and to which degree
shared reference frames are resources that are comparable to shared 
quantum states
\cite{Refefficient,Burgh,GroverRudolph,Enk,GroupCovariantThermo,Decoherence-full}. 
In this article 
we want to understand to what extent timing information 
should be considered
as quantum information by exploring the information loss occurring 
when it is copied. 
In \cite{clock} we have derived lower bounds on 
the loss of timing 
information in terms of Fisher-information for the broadcasting problem.
To our knowledge, no results in terms of
Holevo-information  can be found in the literature.

\section{Broadcasting timing information}

\label{Broad}

Before we pose the problem of broadcasting timing information
(which we have motivated from the time-covariant transmission of 
signals) we 
first state the more general 
problem of  ``broadcasting Holevo-information''.
It is defined
as follows:

\begin{Definition}[Broadcasting Holevo-Information]${}$\\
Given an ensemble $\{p(x),\rho (x)\}_x$ 
of quantum states acting on some Hilbert space $\cH$. Let $\cI$ 
be its Holevo-information. 
Find an optimal broadcasting map in the following sense: 

Let 
$\cH_A$ and $\cH_B$ be some arbitrary additional 
Hilbert spaces and $G$ be a
completely  positive trace-preserving operation 
from the density operators  on $\cH$ on 
the density operators acting on
$\cH_A\otimes \cH_B$.
Let $\cI_A,\cI_B$ denote the 
Holevo-information of the ensembles given by the reduced states
$tr_B(G(\rho_x))$ and $tr_A(G(\rho_x))$, respectively. 

Maximize the average  information
\[
\frac{1}{2}(\cI_A+\cI_B)
\]
over all $\cH_A\otimes \cH_B$ and $G$ such that it gets 
as close to $\cI$ as possible.

We call
\begin{equation}\label{IDiff}
\Delta:=\cI-\frac{1}{2}(\cI_A+\cI_B)
\end{equation}
the broadcasting loss of a given broadcasting operation.
Let $\Delta_{\min}$ be the minimal loss over all broadcasting operations for
a given ensemble. In the context of timing information we will
also use the terminology "$\Delta_{\min}$ of a state $\rho$''
when actually refering to the information loss of the ensemble
 $\{\alpha_t(\rho)\}_{t\in [0,\tau)}$ 
with uniform probability distribution over the whole time period. 
\end{Definition}

\noindent
Due to the monotonicity of the 
Holevo-information of an arbitrary ensemble with
respect to CP maps 
\cite{Referenz}
we  certainly have $\cI_1\leq \cI$ and $\cI_2\leq \cI$. 
It is
natural to conjecture that $\cI=\cI_1=\cI_2$ can only be achieved
if all density matrices commute, which is exactly the case where
usual broadcasting is possible \cite{NoBroadcast}.
It is furthermore obvious that there are maps that provide both parties
with the {\it accessible} information \cite{NC}  
by applying a measurement
to the input state and sending mutually orthogonal quantum states
representing the results to both parties.

Now we apply the definition of broadcasting to an ensemble
given by the
time orbit $(\rho_t)_{t\in [0,\tau]}$ of a dynamical evolution
with period $\tau$ 
with uniform distribution over the whole interval.  
Then the task is to optimally broadcast the timing information 
in the sense of
Definition~\ref{Tim}. Note 
that information {\it differences}
like that one in eq.~(\ref{IDiff})  
maybe well-defined  for systems with continuous spectrum
where the timing information itself is infinite. 
By appropriate limits, one could therefore define the
question on the
information {\it loss} in broadcasting operations also
for systems possessing no time average state. 

To give an impression on the problem of broadcasting 
timing information  we consider the
phase-covariant cloning of an equatorial qubit state, i.e.,
a state
\[
|\psi_t\rangle:= \frac{1}{\sqrt{2}}(|0\rangle+e^{it} |1\rangle) 
\]
with unknown $t\in [0,2\pi)$. In the usual quantum  cloning problem
one tries, for instance, to obtain two  copies whose states
get as close to the original as possible with respect to the
Fidelity.
As shown in \cite{BrussPhase} one can generate two copies 
as mixed states whose Bloch vectors point in the same
direction as that of the original, but are shorter than the original 
by the factor $1/\sqrt{2}$.
Thus, the  
density matrices of the copies have the eigenvalues $1/2 \pm 1/(2\sqrt{2})$.
The Holevo-information of each copy is then given by
the entropy of the time average (which is still one bit) 
minus the above binary entropy when inserting the above eigenvalues:
\begin{eqnarray*}
\cI&=& 1+ \frac{1}{2}(1+\frac{1}{\sqrt{2}}) 
 \log_2 \frac{1}{2}(1+\frac{1}{\sqrt{2}}) 
+\frac{1}{2}(1-\frac{1}{\sqrt{2}}) 
 \log_2 \frac{1}{2}(1-\frac{1}{\sqrt{2}})\\
&\approx& 0.399 \hbox{ bit}\,.
\end{eqnarray*}
The information of the original was $1$.  
Here, even the sum of the amount of information
over both copies is less than the original amount. In other words,
the    average 
information over both copies is even smaller than it was if
we had given one party the original and the other an arbitrary state
that is independent from the input.


\section{Information deficit in pure product states
and entropy power inequalities} 

\label{PureProd}

In the following we will not explicitly consider the
broadcasting operation that generates a bipartite state from
the original. Since this operation can never increase the 
information we focus to the following problem:
Given 
an ensemble 
of bipartite states, compare the Holevo-information of 
the two ensembles $\cI_A$ and $\cI_B$  
defined by the restrictions to the subsystems
to the information $\cI$ of the joint system.
Call $\cI-(\cI_A+\cI_B)/2$ the {\it information deficit}.
 In other words,
the information deficit is the broadcasting loss if  
the broadcasting map is the identity and the original is already 
a bipartite state.

Remarkably, the determination of the deficit 
is non-trivial even when 
the bipartite state  is a product state.
Given the state
\[
|\psi_A\rangle \otimes |\psi_B\rangle \in  \cH_A \otimes \cH_B\,,
\]
where each subsystem is subjected to its own Hamiltonian
$H_A$ and $H_B$, respectively. We may assume without loss of generality
that both Hamiltonians are diagonal and non-degenerate (since
we restrict the attention to the time orbits 
of each state). The distribution of energy values
in the state $|\psi_A\rangle\otimes |\psi_B\rangle$ defines
a joint distribution of two stochastically independent 
classical random variables $X,Y$ by
\begin{equation}\label{energyMeasure}
P(X=x,Y=y):= \langle \psi_A |R_x|\psi_A\rangle
\langle\psi_B|  
Q_y |\psi_B\rangle\,,
\end{equation}
where $R_x$ is defined 
as in Definition~\ref{Tim} and $Q_y$ similarly.
Since $H_A\otimes {\bf 1} +{\bf 1} \otimes H_B$ is 
 the Hamiltonian of the joint system,  its
 timing information is  given by
\begin{equation}\label{JointI}
\cI=S(X+Y) \,,
\end{equation}
where we have decided to use the same symbol for the entropy
of classical random variables as for the von-Neumann entropy of
quantum states. 
The subsystem timing information is
given by 
\begin{equation}\label{SubIn}
\cI_A= S(X) \,\,\,\,\,\hbox{ and } \,\,\,\,\,\cI_B=S(Y)\,.
\end{equation}
Note that it is a well-known problem in classical information theory
to relate the entropy of the distributions of two 
independent random variables
to the entropy of their sum since it addresses the question
how the entropy of a real-valued signal changes when subjected to
an additive noise. 
We rephrase the following result that applies to continuous
distributions.   
For probability densities $P(X)$ 
the continuous entropy is defined by
\[
S(X)=-\int P(x) \ln P(x) dx +c \,,
\]
with an unspecified constant $c$.
For two independent random variables, i.e., when their density 
satisfies $P(x,y)=P(x)P(y)$, we have
the entropy power inequality \cite{Stam}
\[
e^{2 S(X+Y)} \geq e^{2S(X)} + e^{2S(Y)}
\,,
\]
and hence
\begin{eqnarray*}
2 S(X+Y)&\geq & \ln \Big(\frac{1}{2}(2e^{2S(X)}+2e^{2S(Y)})\Big)\\
&\geq& \frac{1}{2} \Big( \ln 2e^{2S(X)} +\ln 2e^{2S(Y)}\Big)\\
&=& \ln 2 + S(X)+S(Y)\,, 
\end{eqnarray*}
where the second inequality follows from the concavity of
the logarithm. Assuming that the spectral measures of $H_A$ and $H_B$ 
are sufficiently distributed over many energy eigenvalues we can
approximate the discrete entropy with the continuous expression for 
appropriate
densities.  After
using eqs.~(\ref{JointI}) and (\ref{SubIn}) we obtain 
\[
\cI \geq \frac{1}{2}(\ln 2 +\cI_A +\cI_B) 
\,.
\]
Note that $\ln 2$ 
corresponds exactly to one bit of information
since the entropy power inequality refers to entropy measured in
 natural units. 
We conclude that for continuous spectrum and product states  the 
timing information of the joint system is at least half a bit more
than the average timing information over both systems.

\section{Information deficit for pure entangled states}

\label{PureEnt}

To estimate the information deficit for entangled states
we will also use  
the joint distribution of $X$ and $Y$ on $\R^2$ given by
\begin{equation}\label{JM}
P(X=x,Y=y):=tr(\rho (R_x\otimes Q_y))\,,
\end{equation}
with the spectral projections $R_x$ and $Q_y$. 
If the bipartite  system is 
in an entangled state, 
eq.~(\ref{SubIn}) is no longer true. 
Moreover, we cannot assume that both Hamiltonians
are ``without loss of generality'' 
non-degenerate since the reduced states may be mixed 
even within a specific  degenerate energy eigenspace.  
However,
eq.~(\ref{JointI}) still holds for pure states. 
We replace eq.~(\ref{SubIn})  by
\[
\cI_A=S(\overline{\rho_A})-S(\rho_A) \,,
\]
where $\rho_A$  
denotes the reduced state on system $A$ and obtain $\cI_B$ in a similar way. 
To derive upper bounds on 
the timing information of the subsystems
we need the following Lemma.

\begin{Lemma}[Average Entropy of Post-Measurement States]${}$\\ 
\label{Ave}
Let $(R_j)_j$ be a complete family of orthogonal projections defining
a measurement
and $\sigma$ be an arbitrary quantum state. Let $S(p)$ be the
Shannon entropy  of the outcome probabilities 
$p_j:=tr(R_j\sigma)$.
Then we have
\[
S(\sum_j R_j \sigma R_j) \leq S(\sigma)+S(p)\,.
\]
\end{Lemma}

\vspace{0.3cm}
\noindent
Proof: 
The statement is equivalent to 
\begin{equation}\label{MesAv}
\sum_j p_j S\Big( \frac{1}{p_j} R_j\sigma R_j \Big) \leq S(\sigma)
\end{equation}
Let $\sigma=\sum_i q_i \sigma_i$ be a decomposition of $\sigma$ into pure states. 
We can consider $S(\sigma)$ as the Holevo-information of the
ensemble $\{q_i,\sigma_i\}_i$.  
Then the left hand side of eq.~(\ref{MesAv}) is equal to
 the Holevo-information 
of the ensemble after the measurement has been applied. 
It can certainly be not greater than the Holevo-information of 
the original ensemble \cite{Referenz}.
$\Box$
\vspace{0.3cm}

\noindent
For our derivation of an upper bound on the information
of the subsystems  
the following Lemma will be crucial.

\begin{Lemma}[Timing Information is  less than Conditional Entropy]
${}$\\\label{RedInf}
Let $\rho$ be a (possibly mixed) state on a bipartite system.
Then the timing information of $A$ and $B$ satisfies
\[
\cI_A \leq S(X|Y), \hspace{1cm} \cI_B \leq S(Y|X) \,,
\]
respectively, 
where the joint distribution of $X$ and $Y$ is defined by eq.~(\ref{JM}). 
\end{Lemma}
\vspace{0.3cm}

\noindent
Proof: By Definition~\ref{Tim} the timing information of system $A$ is given by 
\[
\cI_A=S(\overline{\rho_A})-S(\rho_A)\,. 
\]
We decompose $\rho_A$ into
\[
\rho_A=\sum_y p(y) \rho_{A,y}\,,
\]
where $\rho_{A,y}$ denotes the conditional state given that we had 
measured the energy value $y$ on system $B$. Since $\cI_A$  is the Kullback
Leibler distance between 
$\overline{\rho_A}$ and $\rho_A$  (see \cite{clockentropy}) 
it is convex and we get
\[
\cI_A \leq \sum_y p(y) (S(\overline{\rho_{A,y}})-S(\rho_{A,y}) )\,.
\]
For each specific value $y$ of $Y$ 
\[
S(\overline{\rho_{A,y}})-S(\rho_{A,y})\leq S(X|y) 
\]
holds 
due to Lemma~\ref{Ave}. Taking the convex sum of this inequality
over all $y$  
with weights 
$p(y)$ completes the proof. 
$\Box$

\vspace{0.3cm}
\noindent
Note that there are conditions known \cite{Johnson}, where the 
joint probability
density of two {\it dependent} random variables satisfies the entropy power
inequality 
\[
e^{2S(X+Y)} \geq e^{2S(X|Y)}+e^{2S(Y|X)}\,.
\]
Under such conditions we obtain the same lower bound on the information 
deficit 
as in Section~\ref{PureProd}.

In the general case we have to use other methods to derive
more explicit bounds from
the bounds of Lemma~\ref{RedInf}. For doing so, we will need the 
following lemma.

\begin{Lemma}[Information Deficit and Classical Mutual Information]${}$
\label{QKlassischRed}
The information deficit of a bipartite system being in a pure state 
satisfies  
\[
\Delta \geq \frac{1}{2}\Big(I(X:X+Y) +I(Y:X+Y)\Big)=S(X+Y)-\frac{1}{2}\Big(S(X|Y)+
S(Y|X)\Big)\,,
\]
where $I(.,.)$ denotes the mutual information between classical
random variables \cite{Cover}.
\end{Lemma}

\vspace{0.3cm}

\noindent
Proof: 
Note that the equation $\cI=S(X+Y)$ holds also 
for pure entangled states. 
Using Lemma~\ref{RedInf} we obtain
\begin{eqnarray*}
2\cI- \cI_A -\cI_B &\geq& 2 S(X+Y)-S(X|Y)-S(Y|X) \\
&=&2 S(X+Y)-S(X+Y|Y)-S(X+Y|X)\\
&=&
I(X+Y:Y)+I(X+Y:X)\,.\,\,\,\,\, \Box
\end{eqnarray*}

\vspace{0.3cm}
\noindent
It is possible  to derive bounds on the information loss based on 
Lemma~\ref{QKlassischRed}, since the term on the right hand
vanishes only in the trivial case $S(X+Y)=0$ in which the joint system
contains no timing information at all. To show this we observe that
there is no joint measure where $X$ and $Y$ are both  uncorrelated to
$X+Y$. This is seen from 
\begin{equation}\label{CV}
C(X,X+Y)+C(Y,X+Y)=V(X+Y)\,,
\end{equation}
where $C(.,.)$ denotes the covariance and $V(.)$ the variance.
However, to derive lower bounds on the mutual information 
based on these
covariance terms requires additional assumptions on the distribution.
We will deal with this point later. 

In order to apply the bounds of Lemma~\ref{QKlassischRed} 
it can be convenient to relate them 
to other information-theoretic quantities:

\begin{Lemma}[Mutual Information and Relative Entropy]${}$\\ \label{Con}
Let $X$ and $Y$ be two real-valued random variables and $P$ the
corresponding joint distribution on $\R^2$ with discrete support.
Let $P_{-X}$ and  $P_{X+Y}$ 
denote the marginal distribution for $-X$ and $X+Y$, respectively.  
Denote the
convolution  of both by $P_X *P_{X+Y}$. Then
we have
\begin{equation}\label{PY}
I(X:X+Y) \geq K(P_{Y} || P_{-X} * P_{X+Y})
\end{equation}
and
\begin{equation}\label{PX}
I(Y:X+Y) \geq K(P_{X} || P_{-Y} * P_{X+Y})\,.
\end{equation}
Moreover, we have the symmetrized statement
\begin{eqnarray}\label{PSymm}
&&I(X:X+Y)+I(Y:X+Y) \nonumber \\
&\geq &
 K\Big(\frac{1}{2}(P_X +P_Y)\Big|\Big| \frac{1}{2}(P_{-X}+P_{-Y}) * P_{X+Y}\Big)\,.
\end{eqnarray}
\end{Lemma}

\noindent
Proof: We define measures on $\R^2$ by 
\[
Q(X=a,Y=b):=P(X+Y=a+b)P(Y=b)
\]
and
\[
R(X=a,Y=b):=P(X+Y=a+b)P(X=a)\,.
\]
Then we can rewrite the mutual information on the left hand side
as Kullback-Leibler distances:
\[
I(X:X+Y)=K(P||R)
\]
and
\[
I(Y:X+Y)=K(P||Q)\,.
\]
Due to the monotonicity of relative entropy distance under marginalization
\cite{OhyaPetz} we have
\[
K(P||Q) \geq K(P_X||Q_X)
\]
where $P_X$ and $Q_X$ denote the marginal distribution of $X$ according to
$P$ and $Q$, respectively, i.e., $Q_X(X=a):=Q(X=a)$.
Similarly
\[
K(P||R) \geq K(P_Y||R_Y)\,.
\]
We have
\begin{eqnarray*}
Q(X=a)&=&\sum_b Q(X=a,Y=b)\\&=&\sum_b P(X+Y=a+b)P(Y=b)\\
&=&\sum_c P(X+Y=c) P(Y=c-a)\\ &=&\sum_c P(X+Y=c) P(X=c-a)\,.
\end{eqnarray*}
Hence the marginal distribution $Q_X$ of $Q$ 
is the convolution product $P_{X+Y}* P_{-X}$ and
the marginal distribution $R_Y$ of $R$ is the product 
$P_{X+Y}*P_{-Y}$. This proves inequalities (\ref{PY}) and (\ref{PX}). 

We obtain the symmetrized statement from the convexity of
relative entropy distance \cite{Cover}.
$\Box$ 

\vspace{0.3cm}
\noindent
After applying Lemma~\ref{Con} 
and and Lemma~\ref{QKlassischRed} we conclude:

\begin{Theorem}[Information Deficit for Pure States]${}$\\
\label{HauptE}
Given a  pure state of a bipartite system $A\times B$. 
Let $P_X$, $P_Y$ and $P_{X+Y}$
denote the probability distributions for the energy of $A$, $B$ and 
$A\times B$, respectively. Then the difference between the
joint timing information and the average information of the subsystems
satisfies
\begin{eqnarray*}
\Delta &\geq& K(P_X||P_{-Y}*P_{X+Y})
+ K(P_Y||P_{-X}*P_{X+Y})\\
&\geq & K\Big(\frac{1}{2}(P_X+P_Y)\Big|\Big| 
\frac{1}{2}(P_{-X}+P_{-Y})*P_{X+Y}\Big)\,.
\end{eqnarray*} 
\end{Theorem}

\noindent
The intuitive content of Theorem~\ref{HauptE} is 
the following. If the energy uncertainty of
$A$ and $B$ are both on the same scale
as the uncertainty of  
$X+Y$, the convolution with $P_{X+Y}$ adds a non-negligible amount of 
uncertainty
 to $(P_X+P_{Y})/2$, which implies that the new distribution
obtained by adding additional noise
cannot be close to the original distribution of $X$. 

It is often helpful to consider measures that are symmetric with respect
to exchanging $X$ and $Y$, i.e., $P(X=x,Y=y)=P(X=y,Y=x)$. The following 
Lemma shows that lower bounds on $I(X:X+Y)+I(Y:X+Y)$  
for symmetric joint measures  automatically provide 
bounds for asymmetric measures:

\begin{Lemma}[Symmetrization]${}$\\ \label{Symm}
Let $P$ be a joint distribution of $X$ and $Y$ and 
$\overline{P}$ its symmetrization
$
\overline{P}:=(P+P')/2\,,
$
where 
\[
P'(X=x,Y=y):=P(X=y,Y=x)\,.
\]
Then we have
\[
I_P(X:X+Y) +I_P(Y:X+Y)\geq I_{\overline{P}}(X:X+Y)+ 
I_{\overline{P}}(Y:X+Y)\,,
\]
where $I_P(.,.)$ refers to the mutual information induced by the measure
$P$. 
\end{Lemma}

\vspace{0.3cm}
\noindent
Proof:  
We write
\[
P(X=x,Y=y)=P(X=x|X+Y=x+y)P(X+Y=x+y)\,.
\]
We obtain such a representation also for $P'$ by
replacing only the conditional $P(X|X+Y)$ since the marginal distribution
on $X+Y$ coincides for $P$ and $P'$. 
Then the Lemma follows already from the convexity of mutual information
with respect to convex sums 
of conditionals 
with fixed marginals (Theorem 2.7.3 in \cite{Cover}).
$\Box$.

\vspace{0.3cm}
\noindent
A simple bound on the information deficit can be provided in terms of
the fourth moments of the signal energies:

\begin{Theorem}[Information Deficit in Terms of Energy]${}$\\
\label{E4}
Given a pure bipartite state on $A\times B$. 
Let $(\Delta E)^2$ denote the variance of the total energy
and $\langle E^4_j \rangle$ denote the $4$th moment
of the energy of system $j=A,B$ and $\langle E^4\rangle$ 
be the fourth moment of the total energy.   
Then the 
information  deficit (measured in natural units)
satisfies
\[
\Delta \geq   
 \frac{(\Delta E)^8}{64\,(\langle E_A^4\rangle + \langle E_B^4\rangle)
\langle E^4 \rangle}\,.
\]
\end{Theorem}

\vspace{0.3cm}
\noindent
Proof: Let $P$, as above, 
be the discrete probability measure on $\R^2$ 
describing the energy distribution of the bipartite system.  
We begin by assuming that
$P$ is symmetric (see Lemma \ref{Symm}). Then we have
$C(X,X+Y)=V(X+Y)/2$ (see eq. (\ref{CV})). 
We define a measure  $R$ as in the proof of Theorem \ref{HauptE}
and we can rewrite the covariance as
\[
C(X,X+Y)=\sum_{xy} x(x+y)  (P(x,y)-R(x,y))
\]
With $Z:=X+Y$ we have
\begin{eqnarray*}
\frac{1}{4}
V^2(X+Y)=C(X,Z)^2&=&\Big|\sum_{xz} xz \sqrt{(P(x,z-x)-R(x,z-x))} \\
&\times& \sqrt{(P(x,z-x)-R(x,z-x))}\Big|^2  \\ &\leq& 
\sum_{xz} x^2z^2 |P(x,z-x)-R(x,z-x)| \\
&\times &
\sum_{xz} |P(x,z-x)-R(x,z-x)| \\ &\leq&
\sum_{xz} x^2 z^2 (P(x,z-x) +R(x,z-x))
\|P-R\|_1\\ & =&
(\langle X^2Z^2 \rangle +\langle X^2\rangle \langle Z^2\rangle 
 )\,\|P-R\|_1\\ &\leq&
2 \sqrt{\langle X^4\rangle \langle Z^4\rangle}\, \|P-R\|_1\,.
\end{eqnarray*}
From the first line to the second we have used 
the Cauchy Schwarz inequality which shows also
$\langle X^2Z^2\rangle \leq \sqrt{\langle X^4\rangle \langle 
Z^4\rangle}$ as well as $\langle X^2\rangle \leq 
\sqrt{\langle X^4\rangle }$. 

We recall the bound
\[
K(P||R)\geq \frac{1}{2}\|P-R\|^2
\]
(see Lemma 12.6.1 in \cite{Cover})
for the relative 
entropy measured in natural units.
Then we obtain
\[
\frac{1}{2}\|P-R\|^2_1  \geq
\frac{V(X+Y)^4}{128\,\langle X^4\rangle \langle (X+Y)^4\rangle} \,.  
\]
This implies
\[
I(X:X+Y)\geq 
\frac{V(X+Y)^4}{128\,\langle X^4\rangle \langle (X+Y)^4\rangle}\,.
\]
If we consider an asymmetric measure $P$ we have 
to symmetrize it first. Then we  replace
$\langle X^4\rangle $ with $(\langle X^4\rangle +\langle 
Y^4\rangle) /2$ since the fourth moment of $Y$ 
with respect to the original measure $P$ coincides with the fourth moment
of  $X$ when calculated with respect to the reflected measure
$P'(X=x,Y=y):=P(X=y,Y=x)$. 
Using Lemma~\ref{QKlassischRed} this proves the statement
when replacing the statistical moments of $X, Y$, and $X+Y$  
with 
the more physical terms $\langle E_A^4\rangle$, $\langle E_B^4\rangle$ and
$\langle E^4\rangle$. 
 $\Box$

\section{Quantum capacity required for lossless transmission}

\label{Class}

In this
 section we will derive lower bounds on the quantum capacity
required to transmit an ensemble with some fixed maximal information loss. 
The idea of the argument is the following.
Assume that the timing information of $G(\rho)$ is exactly 
the same as that of $\rho$. 
Assume furthermore that $G$ has zero quantum capacity.
This implies, roughly speaking, that $G$ can be modeled by 
a unitary that copies as much information
to the  environment 
as the amount of 
information that passes the channel. But if this would be 
the case 
we  had perfect broadcast of Holevo-information, an operation
that we consider unlikely to be possible for non-commuting ensembles
like time orbits. To put this argument 
on a solid basis, we rephrase  the following 
result of Devetak \cite{Devetak}. 
Recall that 
the private information capacity (see \cite{Devetak} for a formal definition)
is the maximal number of encoded qubits per transmitted qubits, 
that two parties, the sender  Alice and the receiver Bob, can asymptotically 
achieve in a protocol where 
a potential eavesdropper Eve, 
having access to the full environment of the channel,
gets a vanishing amount of information. 
The following theorem relates
the private  information capacity to the information 
the environment obtains when the channel is represented 
by a unitary acting on the system and an abstract environment
being in a pure state\footnote{One should emphasize that the unitary 
extension gives only upper bounds on the information transfered to the 
environment. Real environments are usually in mixed states and can 
therefore destroy quantum superpositions without receiving information
from the system (see \cite{Studie} for details).}.

\begin{Theorem}[Private Channel Capacity]${}$\\ 
\label{PrCC}
Let $G$ be a quantum channel mapping density operators
acting on $\cH$ to density operators acting on the same space.
Let $\cH_E$ be an additional Hilbert space thought of as the
space of the environment. 
Moreover, let 
$U$ be
a unitary acting on $\cH \otimes \cH_E$ and  
$|\phi\rangle\in \cH_E$ be a state
such that 
\[
G(\rho)=tr_2(U(\rho\otimes |\phi\rangle \langle \phi|)U^\dagger)\,.
\]
Let $\rho_x$ with $x\in \cX$ be some finite family of input states
(sent by Alice with probability $p(x)$) 
and 
\[
\sigma_x:=U(\rho_x \otimes |\phi\rangle \langle \phi|)U^\dagger
\]
be the corresponding joint states of the environment and 
the receiver's
(i.e Bob's) system. 
Denote the restrictions to these subsystems by
$\sigma_x^B$ and $\sigma_x^E$, respectively.
Set
\[
I(X:B):= S( \sum_x p(x) \sigma^B_x) -p(x)\sum_x S(\sigma^B_x)\,.
\]
and $I(X:E)$ similarly. 
Define the single copy private channel capacity by
\[
C_1(G) := \sup \{ I(X:B)-I(X:E) \}\,,
\]
where the supremum is taken over all ensembles $(p(X),\rho_x)$.
Let $G^{\otimes l}$ be the $l$-fold copy of $G$. Then the
private channel capacity is given by 
\[
C_p(G)=\lim_{l\to\infty} \frac{1}{l} C_1(G^{\otimes l})\,.
\]
\end{Theorem}

\noindent
Certainly, 
we have $C_p(G) \geq C_1(G)$. This is seen by transmitting 
independently distributed product states through the copies of channels.
We observe:

\begin{Theorem}[Information Loss in Classical Channels]${}$\\
\label{InfLoss}
Let $\{p(x),\rho_x\}_x$ be an ensemble of quantum states
with Holevo-information
\[
I(X:A)=S\Big(\sum_x p(x) \rho_x\Big) -\sum_x p(x) S(\rho_x)\,.
\]
with minimal broadcasting
 loss $\Delta_{\min}$. Let $G$ be some channel with
\[
I(X:B)= S\Big(\sum_x p(x) G(\rho_x)\Big) -\sum_x p(x) S(G(\rho_x))\,.
\]
Then the private channel capacity of $G$ can be bounded from below by
\[
C_p(G)\geq 2 \Big(\Delta_{\min} - (I(X:A) -I(X:B))\Big)\,.
\]
\end{Theorem}

\noindent
Note that 
$I(X:A) -I(X:B)$ is the information loss caused by the channel because
it is the difference between input and output Holevo-information. 
Given a bound for broadcasting the Holevo-information of the
considerd ensemble, we have a lower bound on the quantum capacity
to transmit them without loss. 

\vspace{0.3cm}
\noindent
Proof (of Theorem~\ref{InfLoss}): 
Given some unitary operation $U$ extending the channel $G$.
We have 
\[
\Delta := I(X:A)-\frac{1}{2}(I(X:B)+I(X:E)) \geq \Delta_{\min}
\]
by definition of $\Delta_{min}$ 
and 
\[
C_p(G)\geq  I(X:B)-I(X:E)\,,
\]
by Theorem~\ref{PrCC}.   
Then
simple calculations yield the stated inequality.
$\Box$ 

\vspace{0.3cm}
\noindent
The theorem shows that for states with non-zero $\Delta_{\min}$
(which is probably every non-stationary state $\rho$)
the covariant lossless transmission requires a channel with non-zero  
quantum  capacity. Instead of deriving lower bounds
on $\Delta_{\min}$, i.e., the minimum over all $\Delta$  
we will use the bound from Theorem~\ref{E4} and only obtain
bounds in terms of the fourth moments of the energy distribution. 
However, using this theorem is not straightforward for the
following reason:
Given some assumptions on 
the energy distribution
of the input and output signals of  a device 
we want to derive lower bounds
on the quantum capacity required to transmit the signal
without information loss.
To this end, we use the unitary extension 
of the CP map formalizing the device because we have only derived
bounds for {\it pure} bipartite states.
However, the usual construction of the unitary extension
uses an abstract environment Hilbert space where no ``environment 
Hamiltonian''
is specified.  And, even worse,
 given that we had  specified an arbitrary ``environment 
Hamiltonian'', the unitary
that models the channel could have lead to arbitrary
energy distributions for system plus environment and we obtained no
useful statements on the fourth moments. 

The following Lemma shows that we can construct the unitary extension 
such that it is energy conserving in the constructed joint system.
We have here considered a 
finite dimensional system for technical reasons.

\begin{Lemma}[Unitary Extension of Covariant Operations]${}$\\
\label{Extension}
Let $G$ be a completely positive trace-preserving map on the set of
$d\times d$ density matrices that 
satisfies the covariance condition~(\ref{TCov}) with respect to
the time evolution  generated by a Hamiltonian $H$ acting on $\C^d$. 

Then there is a (not necessarily finite dimensional) Hilbert space
$\cH_E$,  a densely defined Hamiltonian $H_E$ on $\cH_E$ 
with purely discrete spectrum and
an eigenstate
$|\phi\rangle$ of $H_E$ with eigenvalue $0$ such that the
following condition holds:
 
There
exists
a unitary $U$ on $\C^n \otimes\cH_E$ commuting with
the extended Hamiltonian
$H\otimes {\bf 1} + {\bf 1}\otimes H_E$ 
which satisfies
\[
G(\rho) = tr_2 (U (\rho\otimes |\phi \rangle \langle \phi|) U^\dagger)
\]
for all density matrices $\rho$. 
\end{Lemma}

\vspace{0.3cm}
\noindent
Proof: 
We assume without loss of generality that $H$ 
is
diagonal with respect to the canonical basis.
Let
\begin{equation}\label{Kr} 
G(\rho)=\sum_{j=1}^k A_j \rho A_j^\dagger
\end{equation} 
be the Kraus representation of $G$ (see \cite{Kraus}). 
Define $\Sigma:=\{x-y \, | \, x,y \in {\tt spec(H)}\}$ where 
${\tt spec}(H)$ denotes the spectrum of $H$.
As shown in (eqs.~(14) in \cite{TimeCovariant})  
we can choose the Kraus operators
such that for every $A_j$ there is some real number $\sigma_j \in \Sigma$ 
with
\begin{equation}\label{ComRel}
[H,A_j]= \sigma_j A_j\,.
\end{equation}
In other words, the operator $A_j$ 
implements a shift of energy values by $\sigma_j$ in the sense that it
maps eigenstates of $H$ with eigenvalue 
$\lambda$ 
onto states with energy $\lambda+\sigma_j$. 
The idea is to choose a unitary extension such that the energy shift
caused by $A_j$ is compensated by 
the opposite shift in the environment.
Thus, we define the Hamiltonian $H_E$ of the environment such that
all values in $\Sigma$ occur as spectral gaps in $H_E$. 
Set $\cH_E:=l^2(\Z)^{\otimes k}$ and
\[
H_E = \sum_{j=1}^k \sigma_j M_j\,,
\]
where $M_j$ is the multiplication operator
acting on the $j$th component 
\[
M_j:={\bf 1}^{\otimes j-1} \otimes {\tt diag }(\dots,-1,0,1,\dots) \otimes {\bf 1}^{\otimes k-j}\,.
\]
Let
\[
S_j:={\bf 1}^{\otimes j-1} \otimes S \otimes {\bf 1}^{\otimes k-j}
\]
be the unitary left shift on $l^2(\Z)$ acting on the $j$th tensor component
via $S|n\rangle:=|n-1\rangle$ for each $n\in \Z$. 
Define
\[
U:=\sum_{j=1}^k A_j \otimes S_j\,.
\]
To see that $U$ is indeed unitary we consider basis states
\begin{equation}
|l\rangle \otimes | {\bf z}\rangle\,,
\end{equation}
where $l=0,\dots,d-1$ and ${\bf z}$ is in the $k$th fold cartesian product
$\Z^{\times k}$.  They are all mapped onto unit vectors because
$\sum_j \langle l| A_j A_j^\dagger |l\rangle=1$. 
The images of different basis states are clearly mutually orthogonal
whenever they correspond to different $k$-tuples  ${\bf z}$. 
If they have ${\bf z}$ in common, they are also orthogonal 
since we obtain then the inner product
\begin{eqnarray*}
&&\sum_j \langle l|A_j A_j^\dagger |\tilde{l}\rangle 
\langle z_1,\dots,z_j+1,\dots,z_k| z_1,\dots,z_j+1,\dots,z_k\rangle =\\  
&&\sum_j \langle l|A_j A_j^\dagger |\tilde{l}\rangle   
= \langle l| \tilde{l}\rangle =0\,.
\end{eqnarray*}
To see that $U$ commutes with the total Hamiltonian
$H_T:=H\otimes {\bf 1} + {\bf 1} \otimes H_E$ 
we observe that for every eigenstate $|l\rangle$ of $H$  with eigenvalue
$\lambda_l$ we have 
\[
A_j|l\rangle =|\phi_{l,j}\rangle \,,
\]
where $|\phi_{l,j}\rangle$ is  some state with  
\[
H|\phi_{\l,j}\rangle=(\lambda_l+\sigma_j)|\phi_{l,j}\rangle\,.
\]
We have
\[
(A_j \otimes S_j) (|l\rangle \otimes |{\bf z}\rangle)=
|\phi_{l,j} \rangle \otimes |z_1,\dots,z_j-1,\dots,z_k\rangle \,,
\]
which is also an eigenstate of $H_T$ for the  eigenvalue 
$\lambda + \sum_j \sigma_j z_j$ as 
$|l\rangle \otimes |{\bf z}\rangle$ is.
 That is, $U$ maps energy basis states
onto energy basis states with the same eigenvalues, i.e., it commutes
with $H_T$. 
We can now choose $|\phi\rangle:=|{\bf 0}\rangle$ as the state of the 
environment.
$\Box$

\vspace{0.3cm}
\noindent
Note that the state 
$U(\rho \otimes |\phi\rangle \langle \phi |)U^\dagger$ 
appearing in the extension of Theorem~\ref{Extension}
has the same energy distribution 
with respect to the extended Hamiltonian as $\rho$ has with respect
to the original system Hamiltonian. This implies that the 
distribution of energy values in the joint state
of system plus environment is given in terms of the distribution of 
input and output
energies. Hence we may now apply our bounds on the information deficit
to the problem of transmitting the states when only limited
quantum capacity is available:

\begin{Theorem}[Information Loss and Quantum Capacity]${}$\\
\label{ILC}
Given a covariant 
completely positive map $G$ with private channel capacity $C_p(G)$. 
Then the difference between the timing information of 
input and output satisfies
\[
\cI_{in}  -\cI_{out}  \geq  
 \frac{(\Delta E_{in})^8}{64(9\langle E_{out}^4\rangle + 
8\langle E_{in}^4\rangle)
\langle E_{in}^4 \rangle }
        -\frac{1}{2} C_p(G) \,,
\]
where $(\Delta E_{in})^2$ and 
$\langle E_{in}^4\rangle$ refer to the variance and the fourth
moment of the incoming signal and
similarly, $\langle E_{out}^4\rangle$ denotes the fourth 
moment of the outgoing
signal.
\end{Theorem}

\vspace{0.3cm}
\noindent
Proof: 
Construct a unitary energy conserving 
extension of $G$ according to Lemma~\ref{Extension}.
Let $E_{out}:=X$ denote the energy of the output signal and
$Y$ the energy of the environment. This implies that $E_{in}:=X+Y$ 
is the initial
energy. To get a bound for $\langle Y^4\rangle =\langle 
(E_{in}-E_{out})^4\rangle$ 
we use $|E_{in}-E_{out}| \leq |E_{in}|+|E_{out}|$ and hence
$(E_{in}-E_{out})^4 \leq  8 (E_{in}^4 +E_{out}^4)$.  
Then we obtain the statement using Theorem~\ref{E4}. 
$\Box$


\section{Implications for the Energy Loss}

\label{Sec:Thermo}

In this section we want to explain why we expect the broadcasting problem
to be specific to {\it low-power} devices.
 One reason is, certainly,
that in current technology, information processing  devices are 
not Hamiltonian systems. Since the system is not closed, 
a unitary description of the signal propagation is not justified.
Furthermore, quantum broadcasting gets only relevant when the time inaccuracy
of a clock signal 
is not dominated  by classical noise
of highly mixed density operators. In the latter case, the energy-time
uncertainty is irrelevant. This is in agreement with the results 
in Ref.~\cite{clock} showing 
 (in 
terms of Fisher-information) that quantum 
bounds on broadcasting timing information
get relevant when the
signal energy times the considered timing accuracy is 
on the scale of $\hbar$. 
However, there is also another link between energy consumption
of information processing devices and broadcasting problems that we have
not mentioned before. The idea is that loss of {\it timing information}
inevitably leads to loss of {\it free energy} in covariant devices. 
This is shown in \cite{Referenz}. We describe the relevant results.
 
First,  we  need the notion of {\it passive} devices, i.e.,
devices having no additional energy source apart from the considered
incoming signal. In other words, all energy resources are explicitly
included into the description. 

\begin{Definition}[Passive Device]${}$\\
A  device with quantum input state $\rho$ and output 
$G(\rho)$ is called passive if $G$ is implemented without energy supply,
i.e.,
\[
F(G(\rho)) \leq F(\rho) \hspace{2cm} \forall \rho
\]
where $F(\rho):=tr(\rho H) - kT S(\rho)$ is the free energy of the system
in the state $\rho$ with reference temperature $T$ and Boltzmann
constant $k$. 
\end{Definition}

\noindent
We have shown in \cite{Referenz} 
that covariant passive channels that decrease the timing information 
decrease also the free energy. We rephrase this result formally.

\begin{Theorem}[Loss of Timing Information Implies Free Energy Loss]${}$\label{FreeImpl}
Let $G$ be a completely positive trace-preserving map
describing a covariant 
passive device. The free energy loss 
caused by $G$
can be bounded from below by the loss of timing information:
\[
F(\rho)-F(G(\rho))\geq kT\,\Big( \cI(\rho)-\cI(G(\rho))\Big)\,.
\]
\end{Theorem}

\noindent
This shows that the channel can only be thermodynamically reversible 
if it does not subject the signal to a stochastically fluctuating time delay,
i.e., 
it has to conserve the timing information. The result is less
trivial than it may seem at first sight. The increase
of signal entropy caused by the additional time delay could 
in principle be compensated by an increase of its inner energy such that 
the free energy of the system is conserved. The covariance
condition is indeed required to show \cite{Referenz} 
that the free energy splits up into the following two components
\[
F(\rho)= kT \cI(\rho) +F(\overline{\rho})\,, 
\]
which cannot be converted into each other. 

Together with Theorem~\ref{FreeImpl} we even 
obtain statements of the thermodynamical
irreversibility of the signal transmission:

\begin{Theorem}[Free Energy Loss in Classical Channels]${}$\\ \label{FreeCl}
Let $\rho$ be a quantum state whose timing information
has the broadcasting loss $\Delta_{\min}$.  
Then every channel $G$ satisfies
\[
C_p(G)\geq 2 \Big(\Delta_{\min} - \frac{1}{kT}(F(\rho)-F(G(\rho))\Big)\,.
\] 
In particular, for every channel with capacity $C_p(G)=0$
we have
\[
F(\rho)-F(G(\rho))\geq \frac{2}{kT} \Delta_{\min}\,.
\]
\end{Theorem}

We may combine Theorem~\ref{FreeCl}  and Theorem~\ref{ILC}
and obtain the following result:

\begin{Theorem}[Free Energy Conservation and Quantum Capacity]${}$\\ 
\label{End}
Given a passive covariant device $G$ with private channel capacity
$C_p(G)$. Let $G$ be applied to a pure input state $\rho$.
Then the free energy loss caused by applying $G$ to $\rho$ 
satisfies
\[
F(\rho)-F(G(\rho))  \geq  kT\,\Big(
 \frac{(\Delta E_{in})^8}{64(9\langle E_{out}^4\rangle + 
8\langle E_{in}^4\rangle)
\langle E_{in}^4 \rangle }
        -\frac{1}{2} C_p(G) \Big)\,,
\] 
with $E_{in}$ and $E_{out}$ as in Theorem~\ref{ILC}. 
\end{Theorem}

It would be desirable to find similar results for
mixed states. However, it seems to be hard to provide 
general bounds. Nevertheless, Theorem~\ref{End} shows
why time covariance brings aspects of quantum information theory
into the theory of low-power signal processing. 
In the context of synchronization protocols we have
already described  in \cite{SynchrEntropy} why covariance
gives rise to additional limitations of thermodynamically reversible
information transfer with classical channels.


\section{Conclusions}

We have described a quantum broadcasting problem that arises naturally
in classical low power signal processing. If a time-invariant device
transmits a signal such that the output signal contains 
the same amount of Holevo-information about an absolute  time 
frame as the input  the following two alternatives are possible:
Either the channel has non-zero quantum capacity or it has internally
solved a quantum broadcasting problem and copied the same amount of 
information to its environment. But this is not possible provided that
(as we conjecture) the Holevo-information of non-commuting ensembles
cannot be broadcast without loss. It is therefore likely that
the time-covariant transmission of signals in a way that causes
no stochastic time delay of the signal
requires devices with non-zero quantum capacity. 
But avoiding stochastic time delays is, as we have argued
 a necessary requirement in order to avoid loss of free energy. 
Thus, we have described a link  between quantum information theory and
the theory of classical low-power processing.



\end{document}